%% file: asnm.tex
\documentclass[11pt]{olplainarticle}

\usepackage[margin=1in]{geometry} 
\usepackage{setspace}
\usepackage{enumerate}
\usepackage{natbib}
\usepackage[parfill]{parskip}    
\usepackage{graphicx}
\usepackage{amssymb}
\usepackage{epstopdf}
\usepackage{amsmath}
\usepackage[labelfont=bf]{caption}
\usepackage{pdflscape}
\usepackage{afterpage}
\usepackage{longtable}
\usepackage{geometry}
\usepackage{subcaption}
\usepackage{float}
\usepackage{appendix}

\usepackage{url}
\usepackage{hyperref}
\hypersetup{
        colorlinks=true,
        linkcolor=black,
        urlcolor=black,
        citecolor=black,
        breaklinks=true}
\usepackage{enumitem}

\title{Forecasting Net Migration By Age:\\ The Flow-Difference Approach}

\date{}

\author[1]{Hana \v{S}ev\v{c}\'{i}kov\'{a}} 
\author[2]{James Raymer}
\author[3]{Adrian E. Raftery}
\affil[1]{Center for Studies in Demography and Ecology, University of Washington}
\affil[2]{School of Demography, Australian National University}
\affil[3]{Departments of Statistics and Sociology, University of Washington}

\begin{abstract}
Most population projection models require age-specific information on net migration totals as a key demographic component of population change. Existing methods for predicting future patterns of net migration by age have proven inadequate. The main reason is that methods applied to model net migration are unable to distinguish factors influencing the inflows from those influencing the outflows. 
In this paper, we develop two flow-difference methods to produce age-specific forecasts of net migration for counties in the Washington State. One uses a deterministic approach; the other uses a Bayesian approach and includes measures of uncertainty. Both methods model the age-specific flows of in-migration and out-migration to derive age-specific net migration. By including models for in-migration and out-migration, even in the absence of data on such flows, the resulting net migration predictions are greatly improved over existing methods that only model the net migration totals. The estimation intervals from the Bayesian flow-difference method are found to be well calibrated, while the other approaches do not yield such intervals.  The implications for future county-level population projections in Washington State are shown.     
\end{abstract}


\begin{document}

\flushbottom
\maketitle

\section{Introduction}

To produce accurate population forecasts, age- and sex-specific information on the expected change due to population movements is required. In comparison to fertility and mortality, information on migration is less likely to be available or reliable. Thus, demographers have come up with a range of methods to infer migration patterns (e.g., \citealt{stillwell1991}; \citealt{rogers2010}; \citealt{bijak2011}; \citealt{willekens2019}). The most commonly applied methods are used to estimate and project future patterns of net migration totals, which is meant to represent the difference between in-migration and out-migration. However, as data on in-migration and out-migration are often not available, net migration is often estimated indirectly by using the demographic accounting equation, where net migration is equal to Population at time $t+n$ minus Population at time $t$ minus births plus deaths. This approach is often called the residual method or intercensal component method \citep{SiegelHamilton1952,edmonston2004}. Further, age patterns of net migration are also estimated or assumed, usually based on empirical observations of in-migration or out-migration.  

As there is a good chance that measures of populations, births and deaths have errors,  estimates of net migration from the residual method capture both the difference between in-migration and out-migration, as well as errors in the other components. While some attempts have been made to develop net migration age schedules (\citealt{pittenger1974, pittenger1978}), there is no `standard' age schedule of net migration that can be expected to remain stable over time (\citealt{rogers1990}; \citealt{plane1994}, pp. 190---196). Some patterns may resemble the age patterns of in-migration or out-migration, particularly if an area receives or sends considerably more migrants than found in the opposite direction, but there is also the possibility that a mixture of positive, negative and near zero values occur across age groups. Model age schedules designed for inflows and outflows of migration (e.g., \citealt{RogersCastro1981}) are not guaranteed to perform well in such situations.

In this paper we develop a generic approach for deriving age-specific net migration counts that can be applied to any location, national or subnational. It consists of two steps:
\begin{enumerate}
\item Assuming that total net migration is available or can be estimated for a given location, we split this total into estimates of total in-migration and total out-migration.
\item We distribute each of these totals across age and derive their differences, yielding age-specific net migration.
\end{enumerate}

Two methods for the second step are proposed in this paper, which we call the flow-difference method. The first is based on a deterministic model and the second on a Bayesian model. Both estimate parameters of the multiexponential model migration schedule \citep{RogersCastro1981}, i.e. the Rogers-Castro model, to distribute estimates of total in-migration and total out-migration by age before taking the difference to obtain estimates of net migration by age.

Forecasting net migration by age is needed for a wide array of reasons. Arguably, the most important is to produce local, state or national population forecasts. Other reasons include planning services and infrastructure for populations that are changing in their age compositions and understanding the future effects of migration on populations. With migration becoming an increasingly important component of population change in just about every context, increasing the robustness and reliability of net migration forecasts will improve the information needed for planning and service delivery.  

\section{Background}
\label{sec:background}

Population estimations and projections are often produced using the cohort component projection method. This procedure starts with an age-specific population and projects the size of each age group going forward in time using age-specific survivorship information from life tables. Values of net migration can be added to the survivorship equations to account for change due to migration. Projected births, also taking into account survivorship, are used to replace the first age groups, i.e., 0-1 or 0-4 year olds, over time. 

Migration complicates population estimation and projection models because it represents a movement between two populations. Ideally, populations would be linked together using destination-specific out-migration probabilities and survivorship for each age group ~\citep{rogers1995multiregional} --- but such data are rarely available. Instead, populations are estimated or projected independently from each other using net migration or total flows of in-migration and out-migration. Even when origin-destination flows of migration are available, the sparseness of some flows (especially when disaggregated by age) and the added complexity discourages many users from incorporating them into projection models. 

Despite the relative simplicity of incorporating net migration into cohort component projection models, the values themselves are hard to predict. The reason is that they represent a difference in flows. Changes in net migration totals over time may be due to changes in the inflows, changes in the outflows or both. Also, they do not have consistent age patterns that can be relied upon to distribute estimates or projections of net migration, nor can they be related to populations at risk of movement (\citealt{rogers1990}; \citealt{raymerUN2023}). 

Despite the problems of including age-specific net migration for projecting populations, often users have no choice due to data limitations. In such cases, a common approach is to first estimate net migration totals and then distribute these totals across ages, using a distribution derived from the Rogers-Castro multiexponential function \citep{RogersCastro1981}. This, for example, was the approach used by the United Nations for many countries prior to the 2024 World Population Prospects~\citep{WPP2024}. This approach was found to be problematic for international migration ~\citep{raymerUN2023}, and arguably even more so for internal migration ~\citep{Yu&al2023}, where small locations can receive both retirement in-migration and significant out-migration of working age persons. 

Another issue with population estimation and projection is the inclusion of uncertainty. We know that measures of population, births, deaths and migration have error. We also know that assumptions about fertility, mortality and migration are unlikely to perfectly capture the patterns in five, ten, or twenty years into the future. Since 2015, the official UN population projections for all countries have incorporated uncertainty about fertility and mortality using a Bayesian approach, while in 2024, the UN incorporated uncertainty about total net migration for the first time \citep{WPP2024}. Our approach extends this by allowing the inclusion of uncertainty about the age-specific pattern of net migration.

\section{Methodology}
\label{sec:methodology}

We start with a definition of our model and its components. Throughout this paper we use the subscript $x$ for age group, $i$ for location and $t$ for time period. For a list of all notation, see Table~\ref{tab:notation}. 

The data we use represent populations and net migration totals by five-year age groups (i.e., $x$ = 0-4, 5-9, …, 95+) every ten years from 1950-1960 to 2010-2020 for each of the 39 counties in Washington State. The primary source of this dataset is the University of Wisconsin-Madison portal on age-specific net migration~\citep{UWisc2023}. As described in the data portal, the decadal net migration totals were estimated by a residual method using Census population data and vital statistics on births and deaths. The net migration residuals capture the change caused by both internal and international migration. The last open age group in this primary dataset is 75+ years. For our purposes, we distributed the 75+ year age group into older age groups (i.e., 75-79, 80-84, ..., 95+) such that the absolute net totals decrease exponentially.

\begin{table}[t]
\caption{\label{tab:notation} \small Notation used in this manuscript.} 
\begin{center}
{\scriptsize
\begin{tabular}{l|l}
{\bf Notation} & {\bf Description} \\\hline
$x$, $y$, $K$ & age index and number of age groups with $x,y = \{1, \dots, K\}$ ($y$ is used in sum functions)\\
$i$ & location index, e.g. county, state, country \\ 
$t$, $T$  & time index and number of time points\\
$G_{i,t}$ & total net migration for location $i$ at time $t$ \\
$g_{x,i,t}$ & age-specific net migration for age $x$, location $i$ at time $t$ with $\sum_x g_{x,i,t} = G_{i,t}$\\
$A_{i,t}$ & total in-migration for location $i$ at time $t$ \\
$\iota_{x,i,t}$ & age-specific in-migration for age $x$, location $i$ at time $t$ with $\sum_x \iota_{x,i,t} = A_{i,t}$ \\
$B_{i,t}$ & total out-migration for location $i$ at time $t$ with $B_{i,t} \ge 0$ \\
$o_{x,i,t}$ & age-specific out-migration for age $x$, location $i$ at time $t$ with $\sum_x o_{x,i,t} = B_{i,t}$ \\
$P_{i,t}$, $P_{x,i,t}$ & total and age-specific population, respectively, for location $i$ at time $t$ and age $x$  \\
$\Theta$ & 11 parameter vector $\Theta = (a_1, \alpha_1, a_2, \alpha_2, \mu_2, \lambda_2, a_3, \alpha_3, \mu_3, \lambda_3, c)$ \\
$r_x(\Theta)$ & Rogers-Castro curve (as migration rate)  at age $x$ given $\Theta$ \\
$f_{x,i}$, $F_i$ & age-specific perturbation factors with $F_i = (f_{1,i}, \dots, f_{x,i}, \dots, f_{K,i})$ \\
$\Delta_i$ & vector $\Delta_i = (\Theta_i, F_i)$ \\
$r^*_{x,\cdot,t}(\Delta_i)$ & Rogers-Castro curve $r_x(\Theta_i)$ perturbed by $F_i$ and weighted by population at risk with $\sum_x r^*_{x,\cdot,t}(\Delta_i) = 1$\\
$W$ & index of an aggregated region, e.g. state, country, world \\
$m$ & crude migration rate \\
$R_{i,x}$ & perturbation factors of the Deterministic model \\
$\beta_{0,i}$, $\beta_0$, $\beta_1$ & parameters of the mixed-effects model \\
$v_i$ & variance parameter of the Bayesian FDM \\
$s$ & sex \\
$l$ & trajectory and sample index
\end{tabular}
}

\end{center}
\end{table}

One of the main backbones of our model is the so-called Rogers-Castro curve, $r_x$, modeling the age-specific migration rate, defined as migrants divided by population at risk in the same age group. We use the original specification from~\cite{RogersCastro1981} which is a function of an 11-parameter vector 
$\Theta = (a_1, \alpha_1, a_2, \alpha_2, \mu_2, \lambda_2, a_3, \alpha_3, \mu_3, \lambda_3, c)$:
\begin{eqnarray}
r_{x}(\Theta) & = & a_1 \exp(-\alpha_1 x) + \label{eq:rc-preworking}\\
&& a_2 \exp\{-\alpha_2 (x-\mu_2) -\exp[-\lambda_2 (x-\mu_2)]\} + \label{eq:rc-working}\\
&& a_3 \exp\{-\alpha_3 [x-\mu_3]- \exp[-\lambda_3(x-\mu_3)]\} + \label{eq:rc-retirement}\\
&& c \label{eq:rc-c}
\end{eqnarray}

Migration of children is modeled by Equation~(\ref{eq:rc-preworking}), young adults by~(\ref{eq:rc-working}), and retirees by (\ref{eq:rc-retirement}). If any of these phases are non-existent, the corresponding parameters can be set to zero. Extensions of this model include a post-retirement component (\citealt{Rogerslittle1994}; \citealt{Yeung&al2023}) and a student peak \citep{wilson2010}. However, given that our primary data source includes five-year age groups to only 75+ years, we do not include them in our models.

We use a location- and time-specific version of the Rogers-Castro curve, $r^*_{x,i,t}$, that is perturbed by a vector $F_i$ of age-specific factors and weighted by population of location $i$ at time~$t$, as well as normalized to sum to one across ages. For this purpose, we define a vector of all parameters, $\Delta_i$, which  includes a location-specific  Rogers-Castro vector $\Theta_i$ and a vector of age-specific factors $f_{x,i}$, as

\begin{eqnarray}
\Delta_i  & = &  (\Theta_i, F_i) = (\Theta_i, f_{1,i}, \dots, f_{x,i}, \dots, f_{K,i})
\label{eq:delta}
\end{eqnarray}

Then, $r^*_{x,i,t}$ is defined as 
\begin{eqnarray}
\label{eq:rc-star}
r^*_{x,i,t}(\Delta_i) & = & \frac{r_{x}(\Theta_i) f_{x,i} P_{x,i,t}}{\sum_y r_{y}(\Theta_i) f_{y,i} P_{y,i,t}}  \,\, .
\end{eqnarray}
Note that $\sum_x^K r^*_{x,i,t}(\Delta_i)  = 1$ for all $i$ and $t$.

Our flow-difference models (FDM) are based on the decomposition of the total net migration of location $i$ at time~$t$, $G_{i,t} $, into an in-migration total, $A_{i,t}$, and an out-migration total, $B_{i,t}$: 
\begin{eqnarray}
\label{eq:Gt}
G_{i,t} & = & A_{i,t} - B_{i,t} \,\,.
\end{eqnarray}

The age-specific net migration, $g_{x,i,t}$, our ultimate quantity of interest, is then defined by distributing $A_{i,t}$ and $B_{i,t}$ into ages using $r^*_{x,\cdot,t}$ and taking their difference:
\begin{eqnarray}
\iota_{x,i,t} & = & A_{i,t} \cdot r^*_{x,W,t}(\Delta^{(in)}_i) \label{eq:ixt}\\
o_{x,i,t} & = & B_{i,t} \cdot r^*_{x,i,t}(\Delta^{(out)}_i) \label{eq:oxt}\\
g_{x,i,t} & = & \iota_{x,i,t} - o_{x,i,t} \label{eq:gxt}
\end{eqnarray}
where $\Delta^{(in)}_i = (\Theta^{(in)}_i, F^{(in)}_i)$ and  $\Delta^{(out)}_i = (\Theta^{(out)}_i, F^{(out)}_i)$, respectively, contain parameter vectors of the Rogers-Castro's in-migration and out-migration schedules, respectively, specific to location $i$, as well as in-migration and out-migration specific perturbation factors. 

To compute $r^*_{x,i,t}(\Delta^{(out)}_i)$ (Equation~\ref{eq:rc-star}), the population at risk is the population of the corresponding location, e.g., county, state or country, $P_{x,i,t}$. For in-migration, the population at risk is a population of a larger region, for example the country or the state, if the individual locations are counties, or the world if the locations are countries. Thus, we use the index $W$ in denoting $r^*_{x,W,t}(\Delta^{(in)}_i)$ in Equation~(\ref{eq:ixt}).

In the next two subsections, we present two FDMs to estimate and predict age-specific net migration $g_{x,i,t}$, namely a deterministic model and a Bayesian model. They are both based on the idea of deriving estimates of the $\Delta^{(in)}_i$ and $\Delta^{(out)}_i$ vectors. One of the key differences is that the deterministic model derives a single Rogers-Castro vector $\Theta$, identical for all locations, and estimates the optimal perturbation of this ``model" schedule, namely $F^{(in)}_{i}$ and  $F^{(out)}_{i}$, to match the observed data. The Bayesian model on the other hand, ignores the perturbations and estimates $\Theta^{(in)}_i$ and $\Theta^{(out)}_i$ for each location $i$ using Markov Chain Monte Carlo (MCMC).  
 
As mentioned in the Introduction, the generic approach consists of two steps. The first step decomposes $G_{i,t}$ into $A_{i,t}$ and $B_{i,t}$, followed by a decomposition into $\iota_{x,i,t}$ and $o_{x,i,t}$, yielding $g_{x,i,t}$ in the second step. Here, we review two approaches for the first step: a heuristic decomposition method which is paired with the Deterministic FDM, and a mixed-effects model which is paired with the Bayesian FDM. However, as the two decomposition steps can be viewed as somewhat independent, one can use the heuristic approach for the first step together with the Bayesian FDM for estimating $g_{x,i,t}$, and vice versa, namely using the mixed-effects model to derive $A_{i,t}$ and $B_{i,t}$, followed by the Deterministic FDM to derive $g_{x,i,t}$.

 \subsection{Deterministic FDM}
\label{sec:param-model}

 We start by describing a deterministic approach to estimate and predict age-specific net migration. The methodology extends the work of \cite{raymerUN2023} that was developed for the United Nations Population Division (UNPD) for improving age- and sex-specific net international migration estimates and projections for use in the 2024 World Population Prospects~\citep{WPP2024}. For the model used in this paper, the residual estimates of net migration by age are used as part of the estimation process. For Washington State there are seven historical data points that include age-specific net migration for each county. The UNPD, on the other hand, did not have reliable age-specific data that could be used to estimate net international migration by age for most countries in the world. 
 
In the first step, net migration totals, $G_{i,t}$, are disaggregated into total in-migration $A_{i,t}$ and total out-migration $B_{i,t}$ as

\begin{eqnarray}
A_{i,t} & = & P_{i,t} m + 0.5 G_{i,t} \label{eq:det-I} \\
B_{i,t} & = & P_{i,t} m - 0.5 G_{i,t}\nonumber
\end{eqnarray}

The overall migration rate $m$ is not known and has to be determined by the analyst. As migration is generally considered to be a rare event, $m$ should be relatively small. However, it should not be so small that negative estimates are produced for $A_{i,t}$. Through iteration, we found that choosing the multiplier $m=0.7$ produced no negative values in the estimated counts of out-migration by age. With the net migration totals representing 10-year totals, this equates to roughly seven percent of the population of Washington migrating in or out of counties each year. Note that the estimated counts of in-migration and out-migration are mainly used to infer the age patterns of net migration; they are not meant to be realistic predictions of in-migration or out-migration. 

To estimate age-specific net migration, this method focuses on determination of the vector of perturbation factors, while using a model migration schedule as a base, whose parameter vector will be denoted by $\Theta^M$:

\begin{eqnarray}
\Delta^{(in)}_i & = & (\Theta^M, \dots, R_{x,i}, \dots )\label{eq:determ-delta}\\
\Delta^{(out)}_i & = & (\Theta^M, \dots, \frac{1}{R_{x,i}}, \dots )\nonumber
\end{eqnarray}

Here, $R_{x,i}$ are ratios of in-migration age proportions to each county divided by $r_x(\Theta^M)$, to be determined by the user. Thus, $r^*_{x,i,t}(\Delta^{(in)}_i)$ and $r^*_{x,i,t}(\Delta^{(out)}_i)$, respectively, represent in-migration and out-migration age schedules obtained by perturbing the model migration schedule and normalizing to sum to one.

The model migration schedule, $r_x(\Theta^M)$, represents a seven-parameter version of the Rogers-Castro model, i.e., components defined in Equations~(\ref{eq:rc-preworking}), (\ref{eq:rc-working}) and (\ref{eq:rc-c}). We found the following estimates of $\Theta^M$ to work well:

\begin{align}
a_1	&= 0.01 \label{eq:ThetaM}\\
\alpha_1 &= 0.09 \nonumber\\
a_2 &= 0.05\nonumber\\
\alpha_2 &= 0.077\nonumber\\
\mu_2 &= 16.5\nonumber\\
\lambda_2 &= 0.374\nonumber\\
c &= 0.0003. \nonumber
\end{align}

\begin{figure}[th]
\begin{minipage}{0.48\textwidth}
\begin{center}
\includegraphics[width=0.8\textwidth]{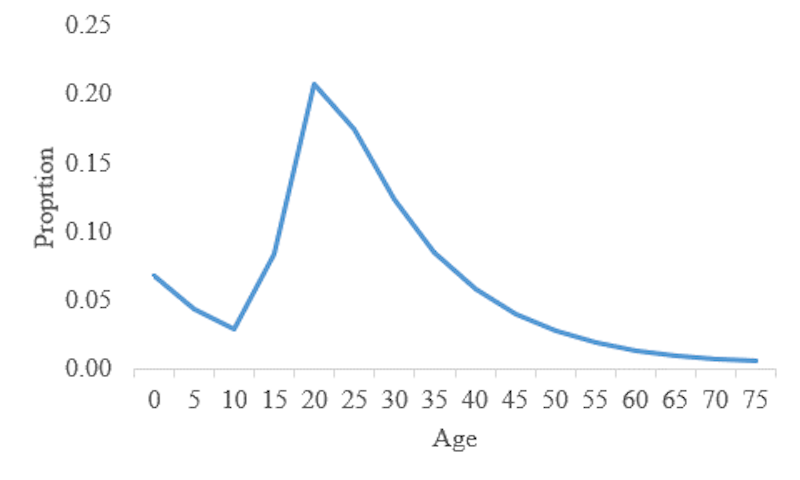}
\end{center}
\end{minipage}
\begin{minipage}{0.48\textwidth}
\begin{center}
\includegraphics[width=0.8\textwidth]{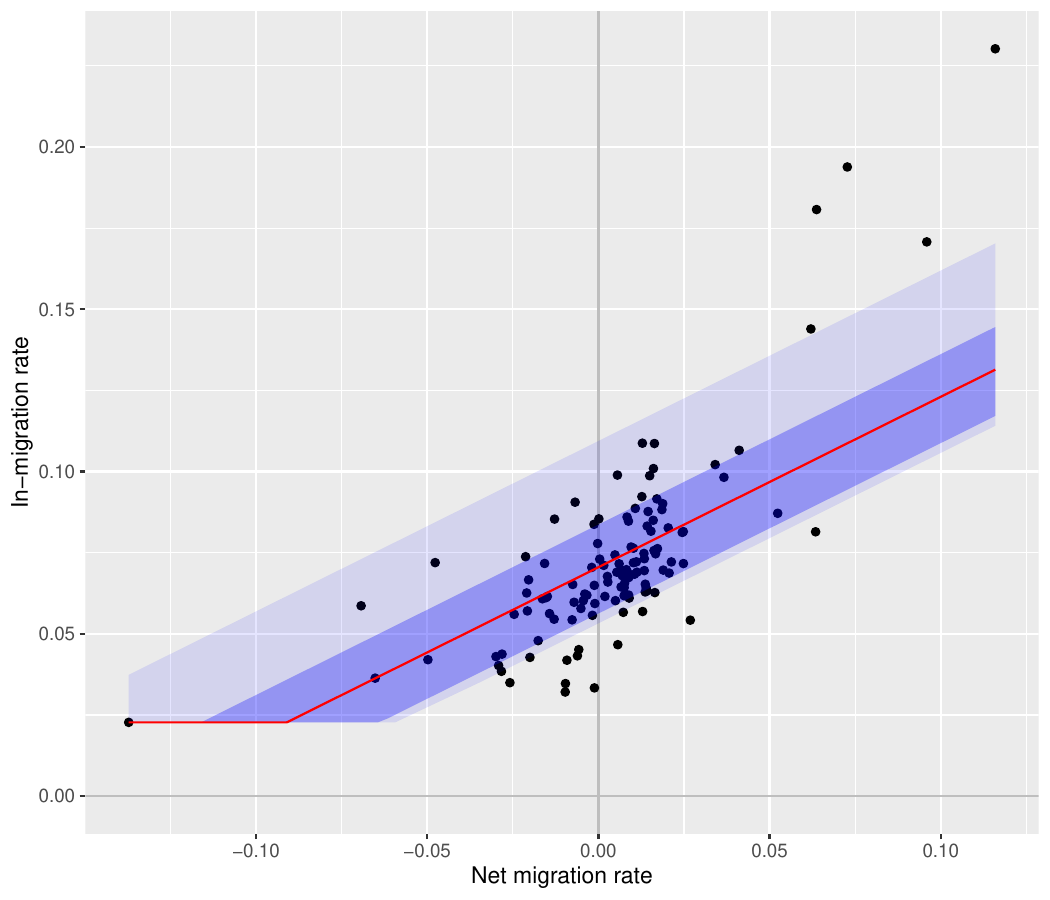}
\end{center}
\end{minipage}
\caption{\label{fig:model-rc-schedule} \label{fig:mem_counties}\small Left panel: Model migration schedule $r_x(\Theta^M)$ with $\Theta^M$ as defined in Equation~(\ref{eq:ThetaM}). Right~panel:~Results from fitting the mixed-effects model (\ref{eq:mixedEffectsModelCounties}) to ACS county-level migration data. The data is shown as black dots. The global mean appears as red line, while the 80 and 95\% intervals are shown as dark and light shaded areas, respectively.}
\end{figure}

The model migration schedule with these parameters is presented in the left panel of Figure~\ref{fig:model-rc-schedule}. Here, we see that the expected overall shape of migration has a downward slope of young children and a large peak of young adults. The propensity to migrate after age 45 is relatively low.

To determine the factors $R_{x,i}$, we proceed as follows. First, to obtain and estimate age-specific in-migration, $\iota_{x,i,t}$, the estimated total in-migration, $A_{i,t}$, is disaggregated by age using the model migration schedule $r_{x}(\Theta^M)$  from Figure~\ref{fig:model-rc-schedule} plus one half of the average observed net migration age pattern, i.e., 

\begin{eqnarray}
\iota_{x,i,t} = A_{i,t}r_{x}(\Theta^M) + 0.5\overline{g}_{x,i} \quad \text{where} \quad \overline{g}_{x,i} = \frac{1}{T}\sum_t g_{x,i,t} \,\, . 
\label{eq:det-ix}
\end{eqnarray}

Then, these age-specific proportions of in-migration are standardized and divided by $r_{x}(\Theta^M)$ to obtain age-specific ratios, $R_{x,i}$, averaged over all available time periods $T$: 

\begin{eqnarray}
R_{x,i} = \frac{1}{T}\sum_t \frac{\iota^*_{x,i,t}}{r_{x}(\Theta^M)} \quad \text{where} \quad \iota^*_{x,i,t} = \iota_{x,i,t} / A_{i,t}
\label{eq:det-Rx}
\end{eqnarray}

We found 1/$R_{x,i}$ to be a good approximation for calculating the out-migration schedule $r^*_{x,i,t}(\Delta^{(out)}_i)$. It also simplified the calculations. 

If the averaging step over $T$ in Equation~(\ref{eq:det-Rx}) is omitted and time-dependent ratios, $R_{x,i,t}$, are used instead of $R_{x,i}$ in Equations~(\ref{eq:determ-delta}), (\ref{eq:oxt}) and~(\ref{eq:gxt}) , the observed net migration totals by age can be recovered. As we observed considerable stability in the ratios over time, the ratios capture much of the deviations from the expected overall migration age profile. Thus, we can use an averaged version of the ratios (Equation~\ref{eq:det-Rx}), averaged over the seven time periods of available data. For going forward in time, one could apply the resulting time-invariant ratios to future population and net migration totals. 
Later in the paper, we will show in detail how $g_{x,i,t}$ can be projected using this method. 

For simplicity, we have omitted to include the population weights used in Equation~(\ref{eq:rc-star}) in the procedure. We believe that for many purposes this is acceptable, for example when the estimates are used for short-term projection, i.e. for time intervals within which the population age structure does not change substantially. 
 
\subsection{Bayesian FDM}
\subsubsection{In- and Out-Migration Totals via Mixed-Effects Model}
\label{sec:AB}
In this section, we present an alternative model to the heuristic deterministic approach in Equation~(\ref{eq:det-I}) for estimating and projecting $A_{i,t}$ and $B_{i,t}$, based on observed data. This model is motivated by \cite{welch2024} who introduced a mixed-effects model to estimate and predict the total in-migration rate, IMR$_{i,t}$, for location $i$ and time $t$ from observed net migration rate, NMR$_{i,t}$. It is defined as follows:

\begin{eqnarray}
\mbox{IMR}_{i,t} &= &\beta_{0,i} + \beta_1 \max \left( \mbox{NMR}_{i,t}, 0 \right) + \varepsilon_{i,t} \,\,,  \label{eq:mixedEffectsModel1}\\
\beta_{0,i} &\sim &\text{Normal}\left(\beta_0, \sigma^2_{between} \right) , \nonumber\\
\varepsilon_{i,t} &\sim & \text{Normal}\left(0, \sigma^2_{within} \right) . \nonumber
\end{eqnarray}
Here, a random intercept term is used to account for the average location-specific
in-migration rate associated with the net migration rate for each time point. 
Location intercepts, $\beta_{0,i}$, are concentrated around a global mean intercept, $\beta_0$.

To estimate the model for the counties of Washington State, we use county-to-county migration flows from the American Community Survey (ACS). Specifically, we use datasets from 2007-2011, 2011-2015, and 2016-2020, each containing estimates of annual immigration, in-migration, and out-migration and for each county. Emigration is estimated using the data in \cite{azose2015}, with an average rate of around $0.3\%$. This is then added to the ACS out-migration counts so that total net migration for each county can be computed. 

We slightly modify the model in Equation~(\ref{eq:mixedEffectsModel1}) which was originally designed for international migration:
\begin{eqnarray}
\mbox{IMR}_{i,t} &= & \max \left( \beta_{0,i} + \beta_1 \mbox{NMR}_{i,t}, \mbox{IMR}_{\min} \right) + \varepsilon_{i,t} \,\, .
\label{eq:mixedEffectsModelCounties}
\end{eqnarray}
The modification allows us to estimate a reasonable in-migration rate even if the net migration is zero or negative. We set IMR$_{\min}$ to the minimum observed IMR$_{i,t}$, namely IMR$_{\min}= 2\%$, suggesting that for negative or zero net migration there is at least 2\% annual in-migration. The model is fitted using the R package {\bf lme4}~\citep{lme42015}. It yields $\beta_1 = 0.52$ and county-specific intercepts $\beta_{0,i}$ ranging from $0.04$ to $0.15$, with global mean $\beta_0 = 0.07$. The right panel of Figure~\ref{fig:mem_counties} shows the data and results from fitting the described mixed-effects model. The estimated random intercepts can be seen in Figure~\ref{fig:mem-beta0-map} in the Appendix.

Translating the results into our framework, while converting rates into counts, given $G_{i,t}$, we derive $A_{i,t}$ and $B_{i,t}$ as

\begin{eqnarray}
A_{i,t} & = & \max\left( \beta_{0,i} P_{i,t} + 0.52  G_{i,t},  0.02P_{i,t}  \right) , \label{eq:AB} \\
B_{i,t}  & = &  A_{i,t} - G_{i,t} \, . \nonumber
\end{eqnarray}

Note that these estimates are on an annual scale. To use them for example on a ten-year scale, we multiply the intercepts and the minimum in-migration rate, $0.02$, by $10$, while the slope, $0.52$, remains the same.

Comparing the main term in Equation~(\ref{eq:AB}), namely $\beta_{0,i} P_{i,t} + 0.52  G_{i,t}$, to Equation~(\ref{eq:det-I}), we can see that the slopes are very similar (around $0.5$). The heuristic equation uses the same intercept rate for all locations ($m$, chosen as $0.07$ annually), while the mixed-effects model estimates location-specific intercept rates $\beta_{0,i}$ with their global mean of $0.07$.

\subsubsection{Age-specific Net Migration}
\label{sec:asnm-bayes}
Our Bayesian approach to solve Equations~(\ref{eq:ixt})-(\ref{eq:gxt}) is to ignore the perturbation vector $F_i$ (or equivalently set it to $F_i = {\bf 1}_K$) and focus on estimating $\Theta^{(in)}_i$ and $\Theta^{(out)}_i$ from the observed data. Thus, we set

\begin{eqnarray}
\Delta^{(in)}_i & = & (\Theta^{(in)}_i, {\bf 1}_K) \,, \label{eq:bayes-delta} \\
\Delta^{(out)}_i & = & (\Theta^{(out)}_i, {\bf 1}_K) \,.\nonumber
\end{eqnarray}

Assuming that $\iota_{x,i,t}$ and $o_{x,i,t}$ in Equations~(\ref{eq:ixt}) and~(\ref{eq:oxt}) are distributed with a probability density function $p(\cdot)$, we can write
\begin{eqnarray}
g_{x,i,t} & \sim & p(\iota_{x,i,t} )  - p(o_{x,i,t} ) \,, \label{eq:pdf}
\end{eqnarray}
with 
\begin{eqnarray}
E[g_{x,i,t}] & = & \iota_{x,i,t} - o_{x,i,t} \, . \label{eq:asnm}
\end{eqnarray}

For example, \cite{Yeung&al2023} assume a Poisson distribution for $p(o_{x,i,t})$. In such a case, if both components are Poisson distributed, their difference could be modeled by
${g_{x,i,t} \sim \text{Skellam}(\iota_{x,i,t}, o_{x,i,t})}$. In our experiments, we found the Skellam distribution problematic in terms of long convergence times with increasing counts, as well as probability intervals being too narrow. 

To avoid these issues, we assume that $p(\iota_{x,i,t} )$ and $p(o_{x,i,t} )$ have negative binomial (NB) distributions. The number of migrants is usually large enough that we can approximate the NB distribution with a Normal distribution, using the fact that if the random variable $X$ has an NB distribution, then Var $(X) = E(X)/v$, where $v$ is a constant parameter \citep{McCullaghNelder1989}.   We can thus approximate their difference with a Normal distribution as follows:

\begin{eqnarray}
g_{x,i,t} & \dot\sim & N(\iota_{x,i,t} - o_{x,i,t}, \,\sigma^2_{x,i,t})\,, \quad \text{with} \nonumber\\
\sigma^2_{x,i,t} & = & \frac{\iota_{x,i,t} + o_{x,i,t}}{v_i} \, . \label{eq:normal}
\end{eqnarray}

Now, since $A_{i,t}$ and $B_{i,t}$ have been estimated, from Equations~(\ref{eq:ixt}), (\ref{eq:oxt}), (\ref{eq:bayes-delta}) and~(\ref{eq:normal}) we see that the parameters to be estimated are vectors $\Theta^{(in)}_i$, $\Theta^{(out)}_i$ and the parameter $v_i$, that is, 23 parameters in total for each location $i$. The model priors are shown in Appendix~\ref{app:priors}.

To avoid identifiability issues, for most counties we include the retirement component in $\Theta^{(in)}_i$ and not in $\Theta^{(out)}_i$ (31 counties), or in $\Theta^{(out)}_i$ and not in $\Theta^{(in)}_i$ (2 counties) or in neither (6 counties). The selection was made considering the historical values of $g_{x,i,t}$. Table~\ref{tab:retirement-comp} in the Appendix gives information about which county is in which category.

\subsection{Estimation Results}
\label{sec:estimation}
The estimation of the Deterministic FDM is straightforward by computing Equations~(\ref{eq:det-I}), (\ref{eq:det-ix}), and~(\ref{eq:det-Rx}). 

For the Bayesian FDM, we first derive $A_{i,t}$ and $B_{i,t}$ using Equation~(\ref{eq:AB}). We estimate the model parameters by Markov chain Monte Carlo using the Stan programing language~\citep{stan}, via the R interface {\bf cmdstanr}~\citep{cmdstanr}. For each county $i$, we ran three Markov Chain Monte Carlo (MCMC) chains, each with $100,000$ iterations after discarding $10,000$ iterations as burn-in.  This yields posterior samples of all 23 parameters, which allows us to construct posterior samples of $r_{x}(\Theta^{(in)}_i)$, $r_{x}(\Theta^{(out)}_i)$ (Equations~\ref{eq:rc-preworking}-\ref{eq:rc-c}),  $r^*_{x,r,t}(\Theta^{(in)}_i)$, $r^*_{x,i,t}(\Theta^{(out)}_i)$ (Equation~\ref{eq:rc-star}), $\iota_{x,i,t}$, $o_{x,i,t}$ (Equations~\ref{eq:ixt}-\ref{eq:oxt}), and finally  $g_{x,i,t}$ (Equation~\ref{eq:normal}). 

\begin{figure}[th]
\begin{center}
\includegraphics[width=0.84\textwidth]{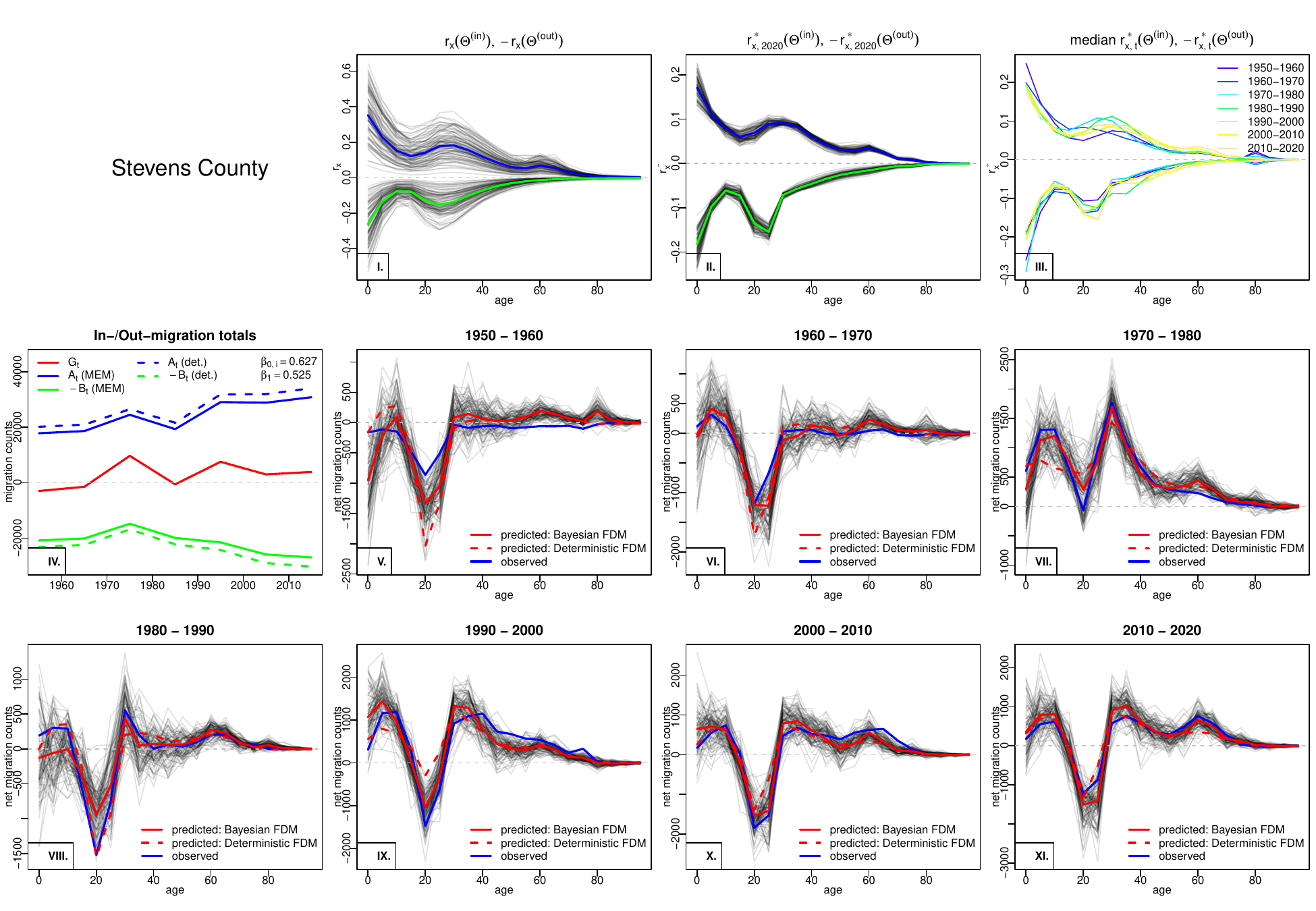}
\end{center}
\caption{\label{fig:stevens-asnmfit} \small Estimation results for Stevens county. The three panels in the first row show the Bayesian posterior distribution of the Rogers-Castro curves as a function of age: Panel I shows the time independent $r_{x}(\Theta^{(in)}_i)$ where the blue line marks its median,  and $-r_{x}(\Theta^{(out)}_i)$ with its median shown as green. Panel II shows the distribution of the population normalized version, $r^*_{x,W,t}(\Theta_i^{(in)})$ and $-r^*_{x,i,t}(\Theta^{(out)}_i)$ for $t=$2010-2020. Panel III shows the medians of $r^*_{x,W,t}(\Theta^{(in)}_i)$ and $-r^*_{x,i,t}(\Theta^{(out)}_i)$ for the seven different values of $t$ used in the estimation. Panel IV in the second row shows the corresponding $G_{i,t}$ (red), $A_{i,t}$ (blue), and $-B_{i,t} = G_{i,t} - A_{i,t}$ (green) as a function of time $t$, including the estimated mixed-effect coefficients used to derive $A_{i,t}$. Results from the deterministic model are shown as dashed lines while the mixed-effects estimates are shown as solid lines. Panels V - XI show for each of the time periods the age-specific data (blue line) and the Bayesian FDM fit (grey lines with its median in solid red). The dashed red lines show the Deterministic FDM fit.
}
\end{figure}

An example of such results, for Stevens county, is shown in Figure~\ref{fig:stevens-asnmfit}. By comparing the red and blue lines in panels V - XI in the second and third rows of the figure, it can be seen that our models predict the observed data well. The Bayesian FDM medians are marked with solid red lines while the Deterministic FDM results are shown as dashed red lines. Also in the figure, in the first row one can see the estimated Bayesian $r_{x}(\Theta^{(in)}_i)$ and  $r_{x}(\Theta^{(out)}_i)$ (panel I.), as well as their population weighted versions $r^*_{x,W,t}(\Theta^{(in)}_i)$ and $r^*_{x,i,t}(\Theta^{(out)}_i)$ (panels II and III). Panel IV in the second row shows results of the decomposition of $G_{i,t}$ into  $A_{i,t}$ and $B_{i,t}$, using the two proposed models. In the next section, we show how these estimates can be used in practical applications, such as in population projections.

\section{Use in Population Projections}
\label{sec:popproj}
\subsection{Method}
\label{sec:proj-method}
One of the main uses of a method for forecasting age-specific net migration is its application in population projections. Here the usual approach is to forecast net migration total which needs to be distributed across sexes and ages before entering the cohort component projection. In the case of probabilistic population projection~\citep{RafteryLi&2012}, one projects a set of trajectories of total net migration rates~\citep{azose2015} which are converted to counts. These again need to be distributed across all sex-age categories. 

Below, we compare three methods: one that is often used in practice and our two FDM approaches. All three methods can be applied either in deterministic or probabilistic projections. The latter only requires one to repeat the procedure for each trajectory of total net migration, yielding probabilistic distribution of age-specific net migration.  For simplicity, we will assume the same migration sex-ratio.

\subsubsection{Basic Rogers-Castro Migration Projection}
Often in practice, one would disaggregate the total net migration into age groups using a basic Rogers-Castro distribution. For a chosen $\Theta$, the age-specific net migration is derived as
\begin{eqnarray}
g_{x,i,t} =  G_{i,t} r_x(\Theta) \,. 
\label{eq:simple-rc}
\end{eqnarray}
We use the same $r_x(\Theta)$ as has been used in the past by the UN for many countries. It can be retrieved using the function {\tt rcastro.schedule(annual=TRUE)} from the {\bf bayesPop} R package~\citep{sevcikova2016bayesPop}. 
We disaggregate this quantity into sexes by dividing it by two, or equivalently, normalizing to $\sum_x r_x(\Theta) = 0.5$ for each sex.

\subsubsection{Deterministic FDM Migration Projection}
For the Deterministic approach, we use $r_x(\Theta^M)$, available by using the vector $\Theta^M$  defined in Equation~(\ref{eq:ThetaM}). As described in the previous sections, using historical data we determine and extract $R_{x,i}$ for each location $i$. Then for each future time $t$ and location $i$, given $G_{i,t}$, we proceed as follows:
\begin{enumerate}
\item For each sex $s$, disaggregate $G^s_{i,t} = G_{i,t}/2$ into $A^s_{i,t}$ and $B^s_{i,t}$ using Equation~(\ref{eq:det-I}) or using the mixed-effects approach (Equation~\ref{eq:AB}). In the rest of the paper we use the former.
\item Considering that population weights were not used, we compute the sex- and age-specific in- and out-migration as 
\begin{eqnarray}
\iota^s_{x,i,t} & = & A^s_{i,t} \frac{r_x(\Theta^M) R_{x,i} }{\sum_y r_y(\Theta^M) R_{y,i} } \label{eq:proj-det-rstar-in}\\
o^s_{x,i,t} & = & B^s_{i,t} \frac{r_x(\Theta^M) / R_{x,i} }{\sum_y r_y(\Theta^M) / R_{y,i} }  \label{eq:proj-det-rstar-out}
\end{eqnarray}
\item The sex- and age-specific net migration is then derived as
\begin{equation}
\label{eq:gt-proj}
g^{s}_{x,i,t}  = \iota^s_{x,i,t} - o^s_{x,i,t} \, .
\end{equation}
\end{enumerate}

\subsubsection{Bayesian FDM Migration Projection}
For the Bayesian approach, we extract a sample from the $r_{x}(\Theta^{(in)}_i)$ and $r_{x}(\Theta^{(out)}_i)$ distribution for each location $i$, estimated as described in the Estimation section. In Figure~\ref{fig:stevens-asnmfit}, these correspond to the absolute values of the grey lines in panel I. For predictive variance, we also extract a sample from the distribution of $v_i$, shown in Equation~(\ref{eq:normal}). For convenience, we select samples of the same size as is our set of trajectories of total net migration counts. We denote the trajectory/sample index by $l$. 

Thus, given $r_{x}(\Theta^{(in)}_{i,l})$, $r_{x}(\Theta^{(out)}_{i,l})$, $v_{i,l}$ and $G_{i,t,l}$, we proceed as follows:
\begin{enumerate}
\item For each sex $s$, disaggregate $G^s_{i,t,l} = G_{i,t,l}/2$ into $A^s_{i,t,l}$ and $B^s_{i,t,l}$. \\
This can be done by applying the estimated random intercept $\beta_{0,i}$ to Equation~(\ref{eq:AB}), or we can also use the deterministic approach in Equation~(\ref{eq:det-I}). Here, we use the former.
\item Using the sex- and age-specific population  at time $t$ of location $i$ for trajectory $l$, $P^s_{x,i,t,l}$, and of an aggregated region $W$ (e.g. state or country), $P^s_{x,W,t,l}$, derive the sex- and age-specific in- and out-migration:
\begin{eqnarray}
\iota^s_{x,i,t,l} & = & A^s_{i,t,l} \frac{r_{x}(\Theta^{(in)}_{i,l})P^s_{x,W,t,l}}{\sum_y r_{y}(\Theta^{(in)}_{i,l}) P^s_{y,W,t,l}} \,, \label{eq:proj-rstar-in}\\
o^s_{x,i,t,l} & = & B^s_{i,t,l}  \frac{r_{x}(\Theta^{(out)}_{i,l})P^s_{x,i,t,l}}{\sum_y r_{y}(\Theta^{(out)}_{i,l}) P^s_{y,i,t,l}} \,.  \label{eq:proj-rstar-out}
\end{eqnarray}
For $P^s_{x,W,t,l}$, to simplify the implementation, we derive the last observation for each $x$ and $s$, and use the same value for all future $t$ and $l$. In the WA counties application, we use WA state as the  aggregated geography for $W$.
\item Draw age-specific net migration from
\begin{eqnarray}
\label{eq:gt-proj}
\tilde{g}^s_{x,i,t,l} & \sim & N\left(\iota^s_{x,i,t,l} - o^s_{x,i,t,l}, \; \sigma^2_{x,i,t,s,l}\right) , \quad \text{with} \label{eq:normalproj}\\
\sigma^2_{x,i,t,s,l} &= &\min\left(\frac{\iota^s_{x,i,t,l} + o^s_{x,i,t,l}}{v_{i,l}}, \left[\frac{P^s_{x,i,t,l}}{2}\right]^2\right)\, .\label{eq:sdproj}
\end{eqnarray}
The second term in~(\ref{eq:sdproj}) decreases the variance if the population is small, in order to avoid a depopulation in the corresponding age group.
\item To ensure that $\sum_x g^s_{x,i,t,l} = G^s_{i,t,l}$, we scale $\tilde{g}^s_{x,i,t,l}$ by
\begin{eqnarray}
g^s_{x,i,t,l} & =  & \tilde{g}^s_{x,i,t,l} - \frac{\sigma_{x,i,t,s,l}}{\sum_y \sigma_{y,i,t,s,l}} \left(\sum_x \tilde{g}^s_{x,i,t,l} + B^s_{i,t,l} - A^s_{i,t,l}\right) \, .\label{eq:gscaled}
\end{eqnarray}
\end{enumerate}

\subsection{Results: Age-Specific Population}
In this section, we compare probabilistic population projections for selected counties in the WA state generated using the two FDM approaches with the basic Rogers-Castro approach used in ~\cite{Yu&al2023}. The three approaches differ only in the way net migration is disaggregated by age, while everything else, including the total net migration rates, remains the same. We use the {\bf bayesPop} R package~\citep{sevcikova2016bayesPop} to perform these three simulations with 1000 trajectories each, generating projections from 2021 to 2050.

\begin{figure}[htb]
\begin{center}
\begin{minipage}{0.32\textwidth}
\includegraphics[width=\textwidth]{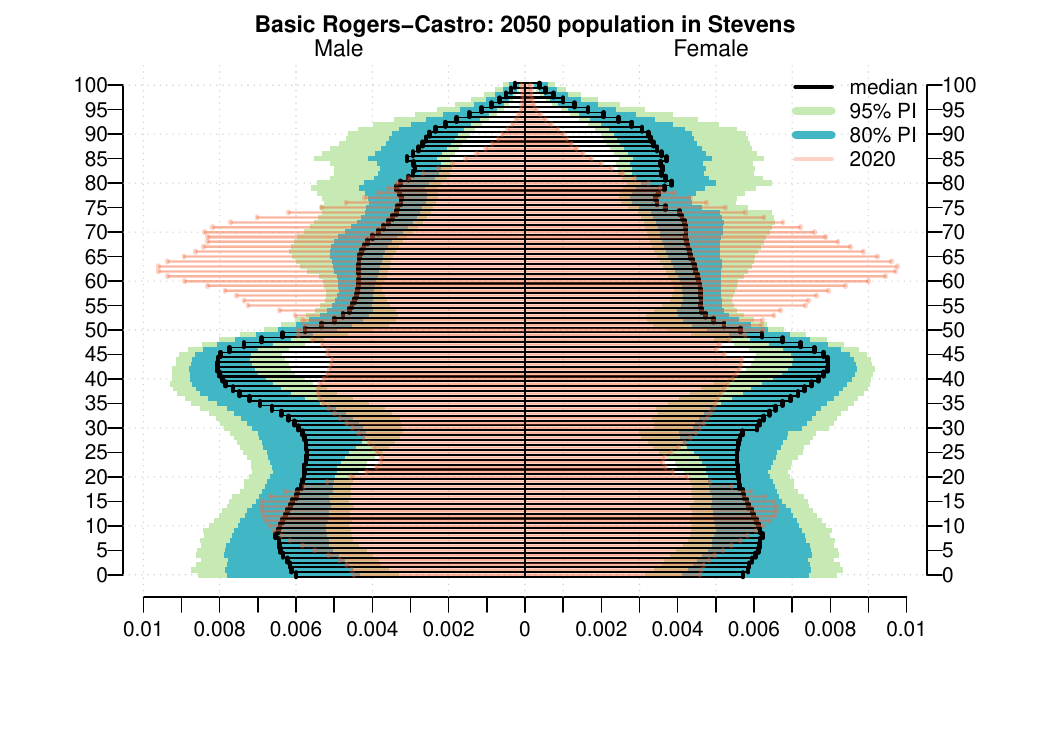}
\end{minipage}
\begin{minipage}{0.32\textwidth}
\includegraphics[width=\textwidth]{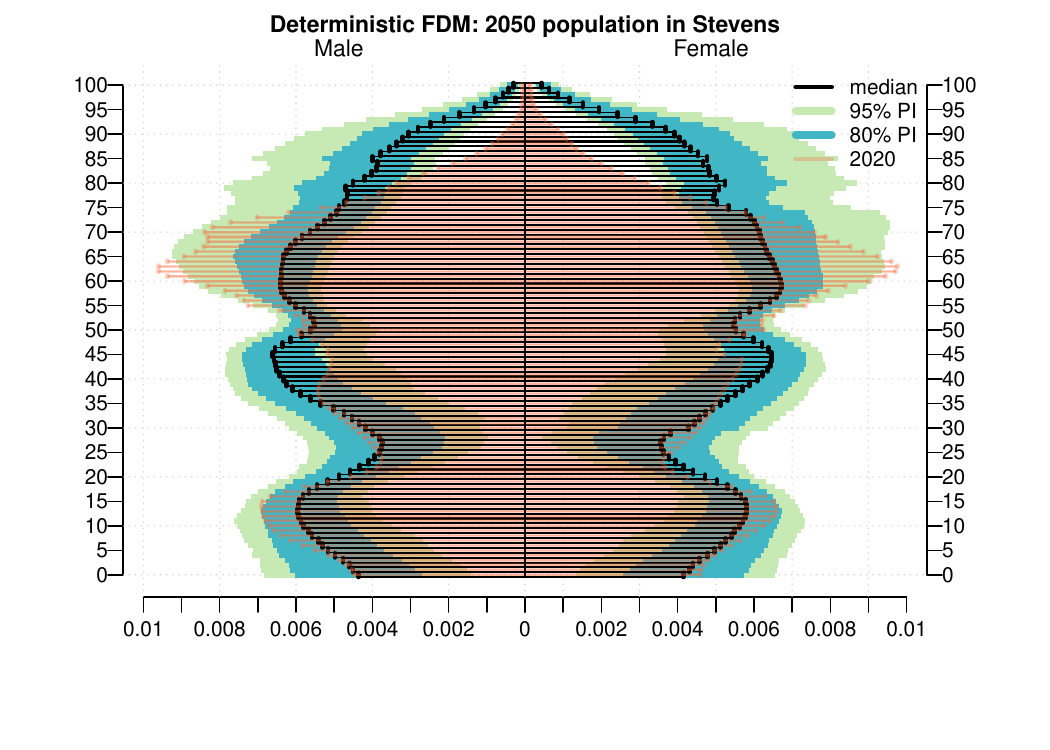}
\end{minipage}
\begin{minipage}{0.32\textwidth}
\includegraphics[width=\textwidth]{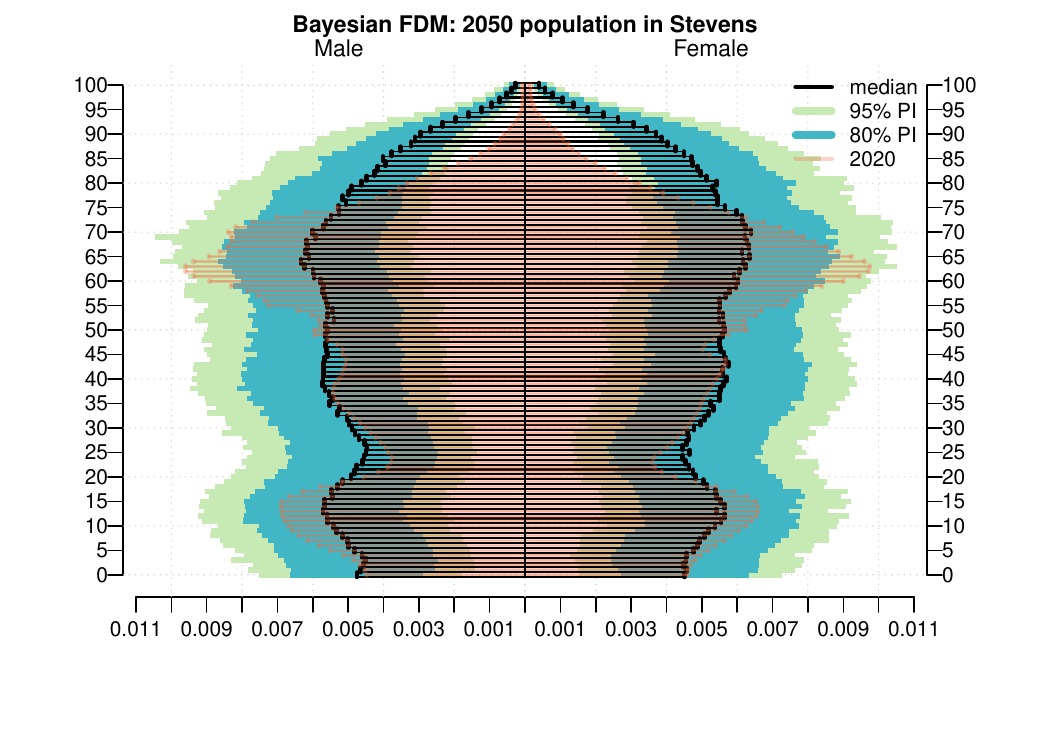}
\end{minipage}

\begin{minipage}{0.32\textwidth}
\includegraphics[width=\textwidth]{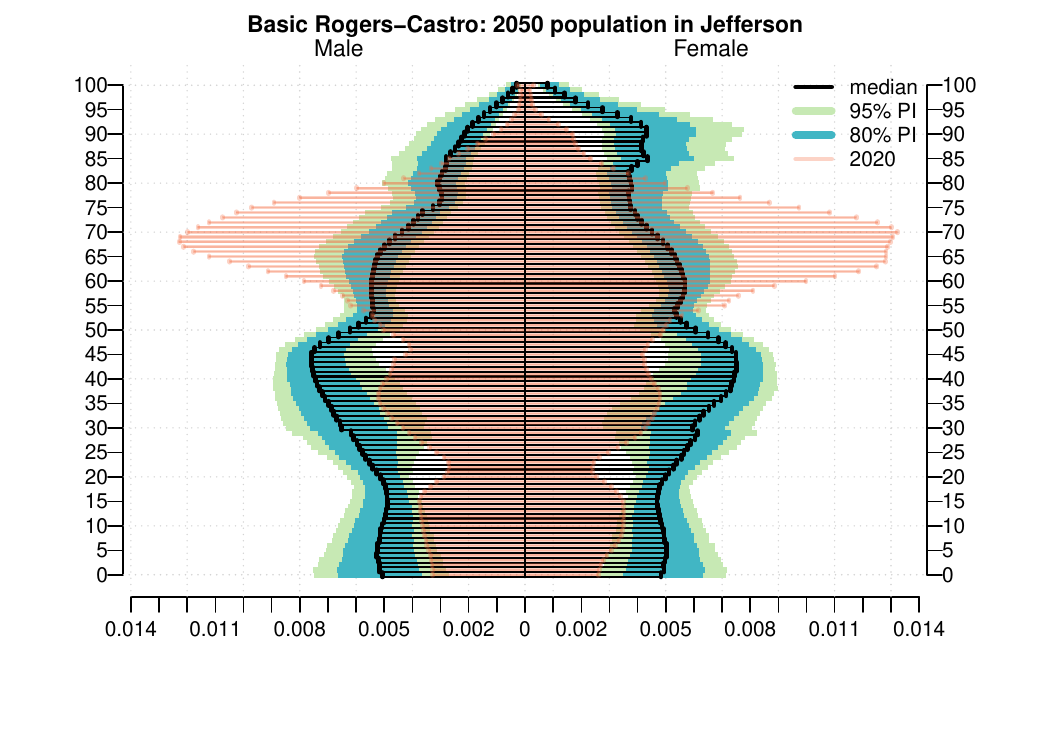}
\end{minipage}
\begin{minipage}{0.32\textwidth}
\includegraphics[width=\textwidth]{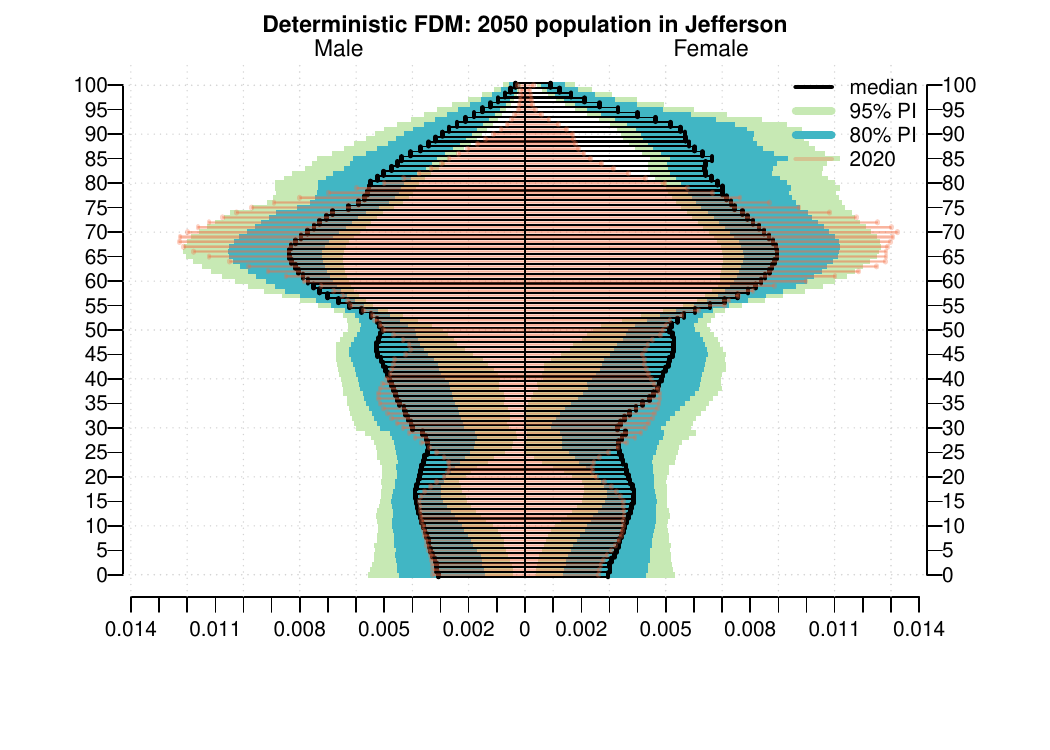}
\end{minipage}
\begin{minipage}{0.32\textwidth}
\includegraphics[width=\textwidth]{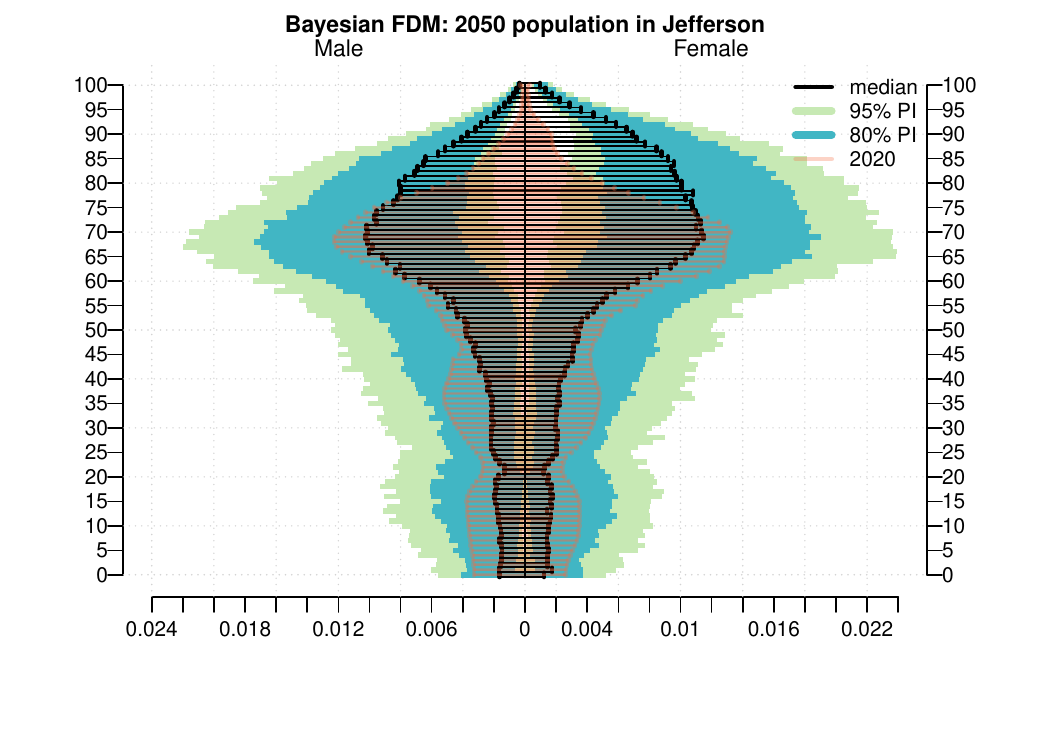}
\end{minipage}
\caption{\label{fig:popproj-by-age} \small 2050 probabilistic population pyramids on a proportional scale for Stevens (top row) and Jefferson (bottom row) counties. The left column shows projections where the total net migration is distributed using a basic Rogers-Castro schedule. The middle and right columns show projections where the Deterministic and Bayesian FDM methods, respectively, were applied. The 2050 median is shown as black line; the 80\% and 95\% probability intervals are shown as blue and green shades. The red bars indicate the last observed (2020) values. }
\end{center}
\end{figure}

How net migration is distributed has a substantial impact on projected populations by age.  Figure~\ref{fig:popproj-by-age} shows probabilistic age-specific population projection in 2050 for two counties, Stevens County (top row) and Jefferson County (bottom row), from the three simulations, on a proportional scale. The left column shows results from the basic Rogers-Castro distribution, the middle column contains results from the Deterministic FDM, while the right column shows the Bayesian FDM results. For reference, each of the population pyramids shows the latest observed data (2020) in red.

Both Stevens County and Jefferson County have historically seen an inflow of people of retirement age, and an outflow of young people. However, the basic Rogers-Castro method ignores these historical patterns and by 2050 the current age distribution is predicted to be completely changed, so that there are more younger than older people.  The FDM approaches on the other hand,  do a good job in preserving the age-specific shapes for most of the ages into the future.  The 95\% probability interval of the FDM methods covers the majority or all of the current bump in old ages.  The source of uncertainty for the basic Rogers-Castro and the Deterministic FDM is the same, namely uncertainty about fertility, mortality and total migration. The Bayesian FDM adds  uncertainty about the migration age distribution, resulting in wider probability intervals.

\subsection{Results: Age-Specific Net Migration}
\begin{figure}[htb]
\includegraphics[width=0.32\textwidth]{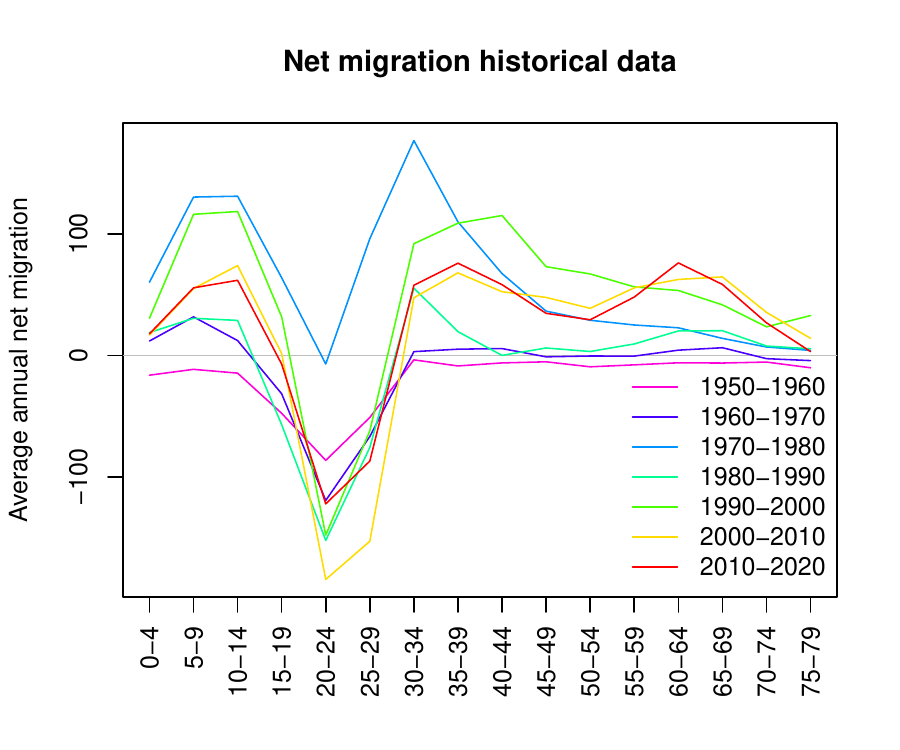}
\includegraphics[width=0.32\textwidth]{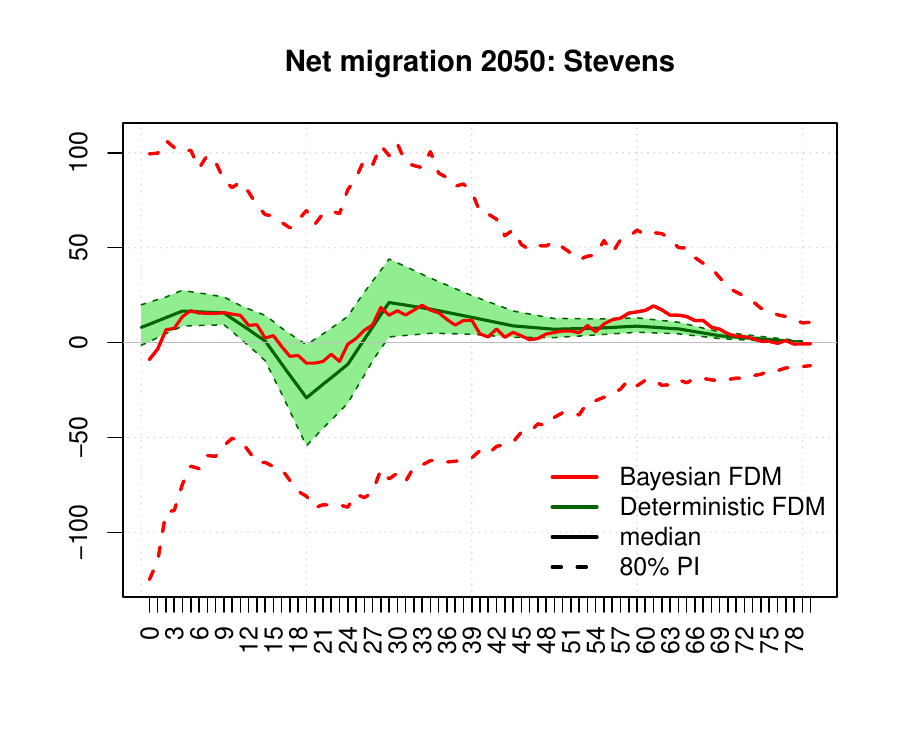}
\includegraphics[width=0.32\textwidth]{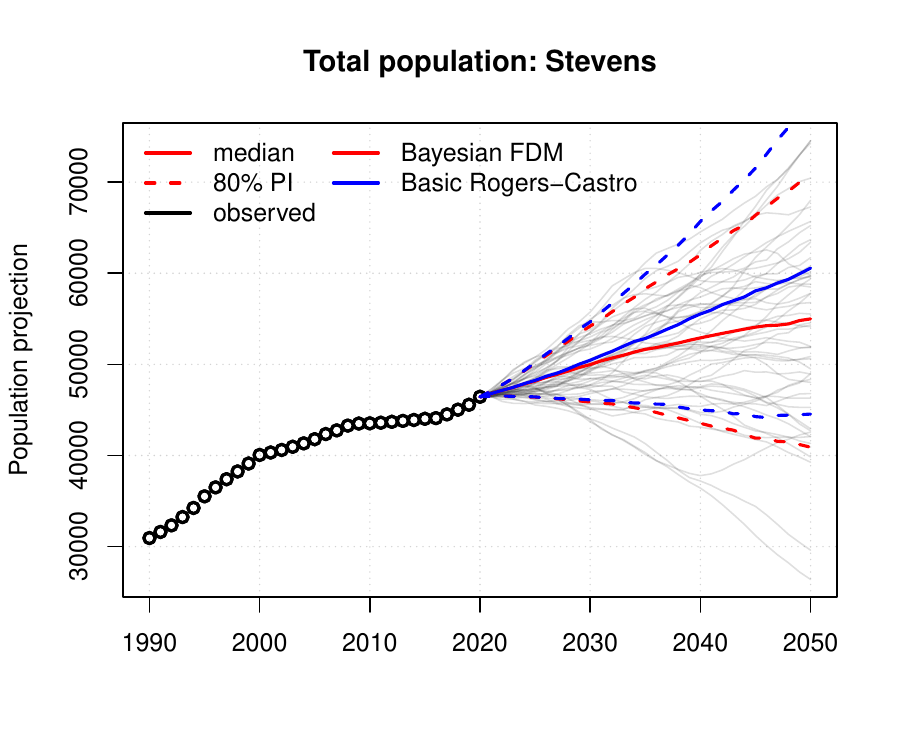}
\caption{\label{fig:migproj-by-age} \small Net migration and total population forecasts for Stevens county. Left panel: historical age-specific net migration (U Wisconsin portal) per 5-year age groups. Middle panel: probabilistic projection of net migration in 2050 for one year age groups, using the Bayesian FDM (red) and the Deterministic FDM (green), with the median lines shown as solid lines, and the 80\% probability interval shown as dashed lines. Right panel: forecasts of total population with the basic Rogers-Castro results shown as blue lines and the Bayesian FDM shown as red lines, while dashed lines being the 80\% probability intervals. }
\end{figure}

The projections of age-specific net migration for Stevens County can be seen in Figure~\ref{fig:migproj-by-age}. The first panel shows the historical data taken from the University of Wisconsin portal, while the second panel shows the projections for 2050 where the two FDM methods are used. In both cases, the projections preserve the historical patterns. In addition, as expected the Bayesian FDM (red curves) yields more uncertainty than the Deterministic FDM (green area).  The impact on total population is shown in the third panel, where for clarity we omitted the Deterministic FDM as it yields similar results to the Bayesian FDM. In the case of Stevens County, projecting migration age patterns that are similar to the observed data reduces the total population in comparison to using the basic Rogers-Castro distribution. This is because the FDM methods project on average positive net migration for older ages and negative net migration for working ages. Note that despite the Bayesian FDM yielding the most uncertainty for age-specific migration and population among the three comparative methods, the amount of uncertainty around the total net migration as well as the total population is the same for all three methods. This is a consequence of the scaling in Equation~(\ref{eq:gscaled}). 

\section{Validation}
\label{sec:results}

\subsection{In-Sample Validation of Net Migration}
\label{sec:validation}
In this section, we assess the three methods from the previous section by performing in-sample validations. As ground truth,
we use the observed age-specific net migration data from the seven time periods as described in the Methodology section and that are shown for Stevens County as blue lines in panels V - XI in Figure~\ref{fig:stevens-asnmfit}. We compare these data to age-specific net migration derived with the three methods for these seven time periods. As comparisons on the scale of counts are dominated by the performance of the models in big counties, we also examine the validation on the scale of net migrants per 100 people. 

We calculate the mean absolute error (MAE), the root mean squared error (RMSE), the bias and the 80\% and 95\% coverages (cov80, cov95). The coverage of a probability interval is defined as the proportion of the time that the true value lies in the interval. Ideally, coverage should be close to the nominal level. Thus, for example, coverage of the 95\% interval would contain close to 95\% of the observed values. For the MAE and RMSE, the smaller the value the better. For the bias, the smaller the absolute value the better.

Table~\ref{tab:validation} shows these quantities averaged over all 20 five-year age groups, 39 counties and seven time periods, that is, over 5,460 data points. The best values are marked in bold. Both FDM methods perform better than the basic Rogers-Castro on MAE and RMSE, while the Bayesian FDM performs best with respect to these measures. The coverage results suggest that the Bayesian FDM method is well calibrated. Since the deterministic and the Rogers-Castro methods are not probabilistic, they do not yield any coverage.

\begin{table}[htb]
\caption{\label{tab:validation} \small In-sample validation of age-specific net migration projection on the scale of counts and rates (migrants per 100 people). The three methods were validated over seven time periods (between 1950-1960  and 2010-2020), 39 counties and 20 5-year age groups, i.e., 5460 values. MAE is mean absolute error. RMSE is root mean squared error. The cov80 and cov95 columns refer to the percentage of the observations that fell within their probability interval.}
\begin{center}
\begin{tabular}{l|l|rrrrr}\hline
\bf Scale & \bf Method & \bf MAE & \bf RMSE  & \bf Bias & \bf cov80 & \bf cov95\\\hline
Counts & Basic Rogers-Castro &     585.4 & 2054.0 & \bf 0.00 & --- & --- \\
& Deterministic FDM & 349.2 & 1207.3 & -0.04 & --- & --- \\
& Bayesian FDM &  \bf 319.7 & \bf 1118.3 & 1.00 & \bf 86.7 & \bf 95.6 \\\hline
Rates & Basic Rogers-Castro &     12.14 & 24.91 & 1.61 & --- & --- \\
& Deterministic FDM & 8.39 & 16.32 & \bf -0.18 & --- & --- \\
& Bayesian FDM &  \bf 7.76 & \bf 13.83 & -1.03 & \bf 86.7 & \bf 95.6 \\\hline
\end{tabular}
\end{center}
\end{table}

\begin{figure}[th]
\begin{center}
\includegraphics[width=0.8\textwidth]{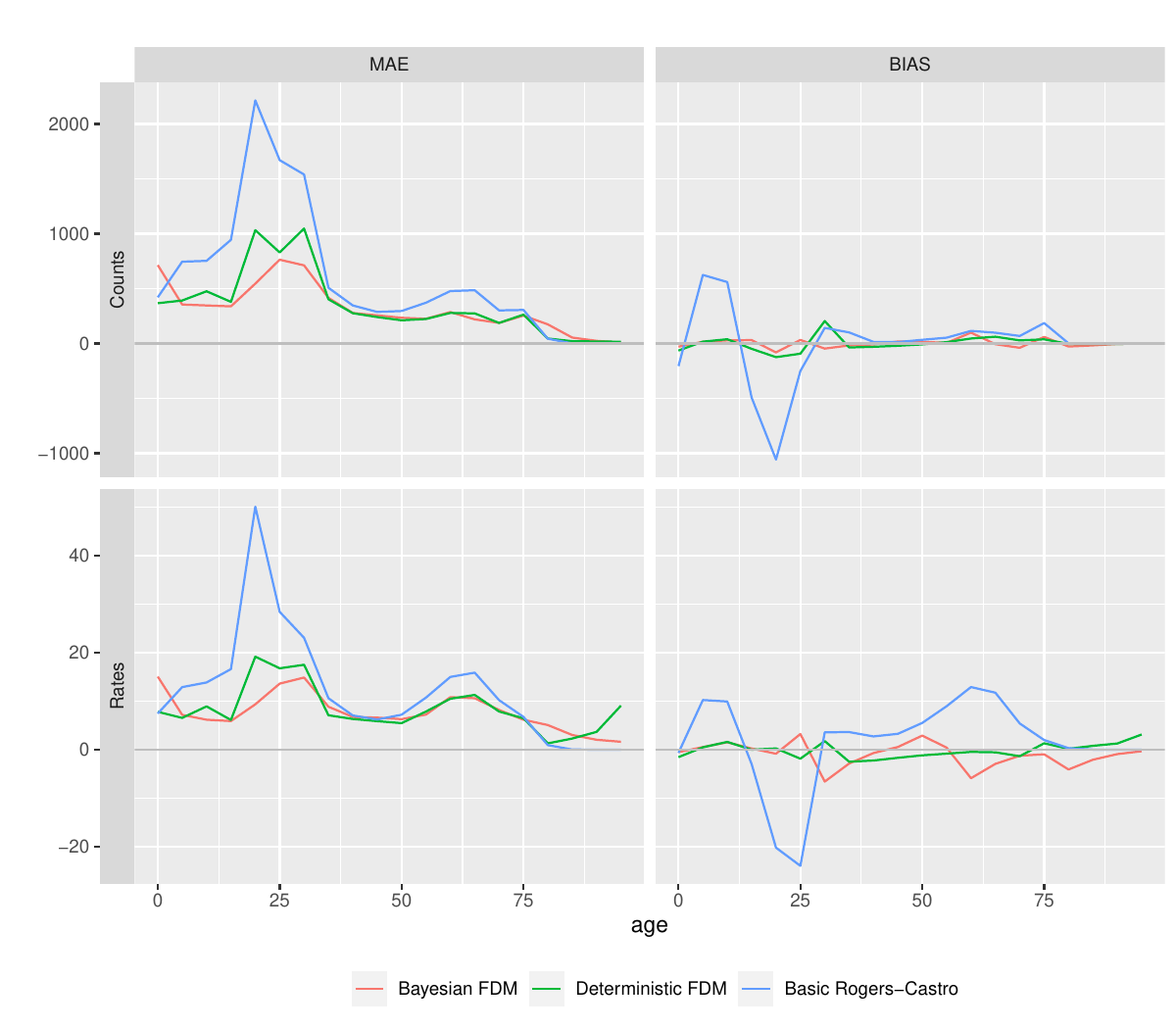}
\end{center}
\caption{\label{fig:validation-by-age} \small In-sample validation of age-specific net migration projection by age for the Bayesian FDM, the Deterministic FDM and a basic Rogers-Castro disaggregation. The first and second row, respectively, shows validation on the scale of counts and rates (migrants per 100 people), respectively. The mean absolute error is shown in the left column, while the bias is shown in the right column. The three methods were validated over seven time periods (between 1950-1960  and 2010-2020) and 39 counties, i.e. each point is an average over 780 values.}
\end{figure}

Validation across ages is explored in Figure~\ref{fig:validation-by-age}. Although the Rogers-Castro method yields a comparable bias on average (as shown in Table~\ref{tab:validation}),  it can be seen in the right column of Figure~\ref{fig:validation-by-age} that it has significant biases for young ages and performs the worst  for old ages as well. The same message applies to the MAE panels on the left.
 
\subsection{Out-of-Sample Validation of Age-Specific Population}
For out-of-sample validation, we examine the predictive performance of the models when applied to project age-specific populations for the 2011-2020 period. Again, we compare our two flow-difference models with the one that uses a Rogers-Castro schedule to disaggregate net migration totals.

For this purpose, we remove any data that enters the cohort component method that is newer than 2010. This also includes re-estimating the TFR and $e_0$ national and subnational models, as well as the subnational total net migration where post-2010 data were removed, as was done in the validation exercise in~\cite{Yu&al2023}. For the FDM methods, we remove age-specific data from 2010-2020 and re-estimate both methods for all counties. Then we produce population projections for 2011-2020. We then compare the age-specific projections for year 2020 with observed data. For the purpose of the validation, the one year age groups on which the framework operates were aggregated into five-year age groups, namely from 0-4 to 100+.

\begin{table}[htb]
\caption{\label{tab:validation-agepop} \small Out-of-sample validation of age-specific population projection generated from 2011 to 2020. The deterministic and Bayesian methods were validated for 2020, over 39 counties and 21 five-year age groups, i.e., 819 values. MAE is mean absolute error. RMSE is root mean squared error. The cov80 and cov95 columns refer to the percentage of the observations that fell within their probability interval.}
\begin{center}
\begin{tabular}{l|l|rrrrr}\hline
\bf Scale & \bf Method & \bf MAE & \bf RMSE & \bf Bias & \bf cov80 & \bf cov95\\\hline
Counts & Basic Rogers-Castro &    847 & 3948 & \bf 95 & 34.1 & 45.3 \\
& Deterministic FDM & 693 & 3219 & 117 & 43.1 & 57.3 \\
& Bayesian FDM & \bf 651 & \bf 2902 & 114 &  \bf 62.6 & \bf 75.5   \\\hline
Shares & Basic Rogers-Castro &     0.63 & 0.98 & 0.0011 &  31.7 & 46.5 \\
& Deterministic FDM & 0.46 & 0.82 & \bf 0.0008 &  37.4 & 56.0 \\
& Bayesian FDM &  \bf 0.44 & \bf 0.71 & 0.0097 & \bf 60.1 & \bf 76.7 \\\hline
\end{tabular}
\end{center}
\end{table}

Table~\ref{tab:validation-agepop} shows the validation results on the scale of counts as well as shares (defined as  the percentage of people in each age group within the total population). The table includes the same validation measures as before, namely the bias, MAE, RMSE and the 80\% and 95\% coverage.  The results paint a similar picture as the in-sample validation, namely that on the MAE and RMSE, both FDM methods perform better than the basic Rogers-Castro, with the Bayesian FDM slightly outperforming the deterministic version. While the Rogers-Castro method does best on the bias on the scale of counts averaged over ages, a further examination of its performance by age reveals that, similarly to Figure~\ref{fig:validation-by-age}, both FDM methods outperform the basic Rogers-Castro method in the labor force and retirement ages (figure not included). Intuitively, these are the ages most impacted by the improved age-specific net migration models.

Since the population projection is probabilistic, all methods yield coverage measures, despite the deterministic nature of two of them. As mentioned above, the uncertainty in these cases is attributed to the uncertainty about total fertility, mortality, and total migration. The Bayesian FDM yields a significantly higher coverage, as the uncertainty around the migration age-distribution is added. However, it can be seen in the coverage columns of Table~\ref{tab:validation-agepop}, that there is more uncertainty in the population projections to be accounted for. Further research on accounting for uncertainty about age-specific fertility and mortality  could improve the age-specific coverage. However, the Bayesian FDM improves the coverage for more aggregated age groups. For example, with three age groups (0-14, 15-64, 65+), the  coverage of the 80\% interval contains $80.3\%$  of the observed population counts.

\section{Conclusion}
\label{sec:discussion}
We have developed two models to distribute net migration totals by age for use in forecasting net migration and inclusion in cohort component projection models. These models were motivated by the methods used to distribute net international migration projections for the UN's World Population Prospects 2024 Revision (\citealt{raymerUN2023}; \citealt{WPP2024}), for subnational migration forecasts in Washington State \citep{Yu&al2023}, and for accounting for age distribution in forecasting international migration \citep{welch2024}. The proposed approaches here extend the previous approaches by generalizing the models to any type of geographic area, provided that some age-specific net migration data are available to inform the forecasts. For illustration, we used the models to produce improved forecasts of net migration age patterns for counties in Washington State.  In doing so, we showed how historical information on age-specific net internal migration can be used to inform forecasts of age-specific net internal migration.  

Both the deterministic and Bayesian flow-difference methods presented in this paper focus on estimating the age patterns of in-migration and out-migration to derive age-specific net migration estimates. This extends the Rogers-Castro curve so that it can be properly applied to net migration. Applying the original Rogers-Castro curve to net migrants does not allow for the possibility that some age groups may exhibit negative values, while other age groups exhibit positive values --- a pattern commonly observed in the data. Further, by doing so, it may distort the projected population age compositions in ways that are unrealistic. While not shown here, the deterministic method and Bayesian method can be applied to further estimate sex-specific patterns or different population groups, provided that some information is available on their historical net migration patterns.


\input{asnm.bbl}
\pagebreak 

\appendix
\appendixpage

\setcounter{figure}{0}
\renewcommand\thefigure{\thesection\arabic{figure}}  

\setcounter{table}{0}
\renewcommand\thetable{\thesection\arabic{table}}

\section{Model Priors and Settings}
\label{app:priors}

The model priors  for the  Rogers-Castro parameters described in the section on Bayesian~FDM, are based on the analysis of almost 200 migration schedules found in Rogers \& Castro (1981), as well as priors used in the {\bf rcbayes} R package~\citep{rcbayes}. We use the same priors  for $\Theta^{(in)}_i$ and $\Theta^{(out)}_i$ for all locations $i$. Therefore for brevity, we leave out the superscripts $(in)$ and $(out)$ as well as the index $i$.

\begin{eqnarray*}
a_1 &\sim & N_{[0,1]}(0, 0.3^2) \\
\alpha_1&\sim & N_{[0,1]}(0,1) \\
a_2 & \sim & N_{[a_1,1]}(0, 0.3^2) \\
\alpha_2&\sim & N_{[0,1]}(0,1) \\
\mu_2&\sim & N_{[0,55]}(25,2^2) \\
\lambda_2&\sim & N_{[\alpha_2,2]}(0,1) \\
a_3 &\sim & N_{[0,1]}(0,0.3^2) \\
\alpha_3&\sim & N_{[0,1]}(0,1) \\
\mu_3&\sim & N_{[55,70]}(63,2^2) \\
\lambda_3&\sim & N_{[0,2]}(0,1) \\
c & \sim & N_{[0,0.01]}(0,0.005^2)\\
v &\sim & U(0,1) \\
\end{eqnarray*}

There are two restrictions on the parameters set to avoid unrealistic migration schedules. First, the constraint for $\lambda_2$, namely $\lambda_2 > \alpha_2$ ensures that the labor part of the Rogers-Castro curve is right-skewed.  Second, the constraint $a_2 > a_1$ assures that the pace of child migration is smaller than that of labor migration. 

\begin{table}
\caption{\label{tab:retirement-comp}\small List of counties. Columns IN and OUT, respectively, provide information if the retirement component of the Rogers-Castro function (Equation~\ref{eq:rc-retirement}) is included in the estimation of the inflow ($r_x(\Theta^{(in)}_i)$) and outflow ($r_x(\Theta^{(out)}_i)$) functions, respectively.}
{\small 
\input wa_counties_ret_comp.tex }
\end{table}

To avoid identifiability issues in the estimation, for most counties we include the retirement component in $\Theta^{(in)}_i$ and not in $\Theta^{(out)}_i$, or in $\Theta^{(out)}_i$ and not in $\Theta^{(in)}_i$ or in neither. Table~\ref{tab:retirement-comp} gives information about which county is in which category.

\section{Mixed-Effects Model -- Results}
\label{app:mem}

\begin{figure}[th]
\begin{center}
\includegraphics[width=0.9\textwidth]{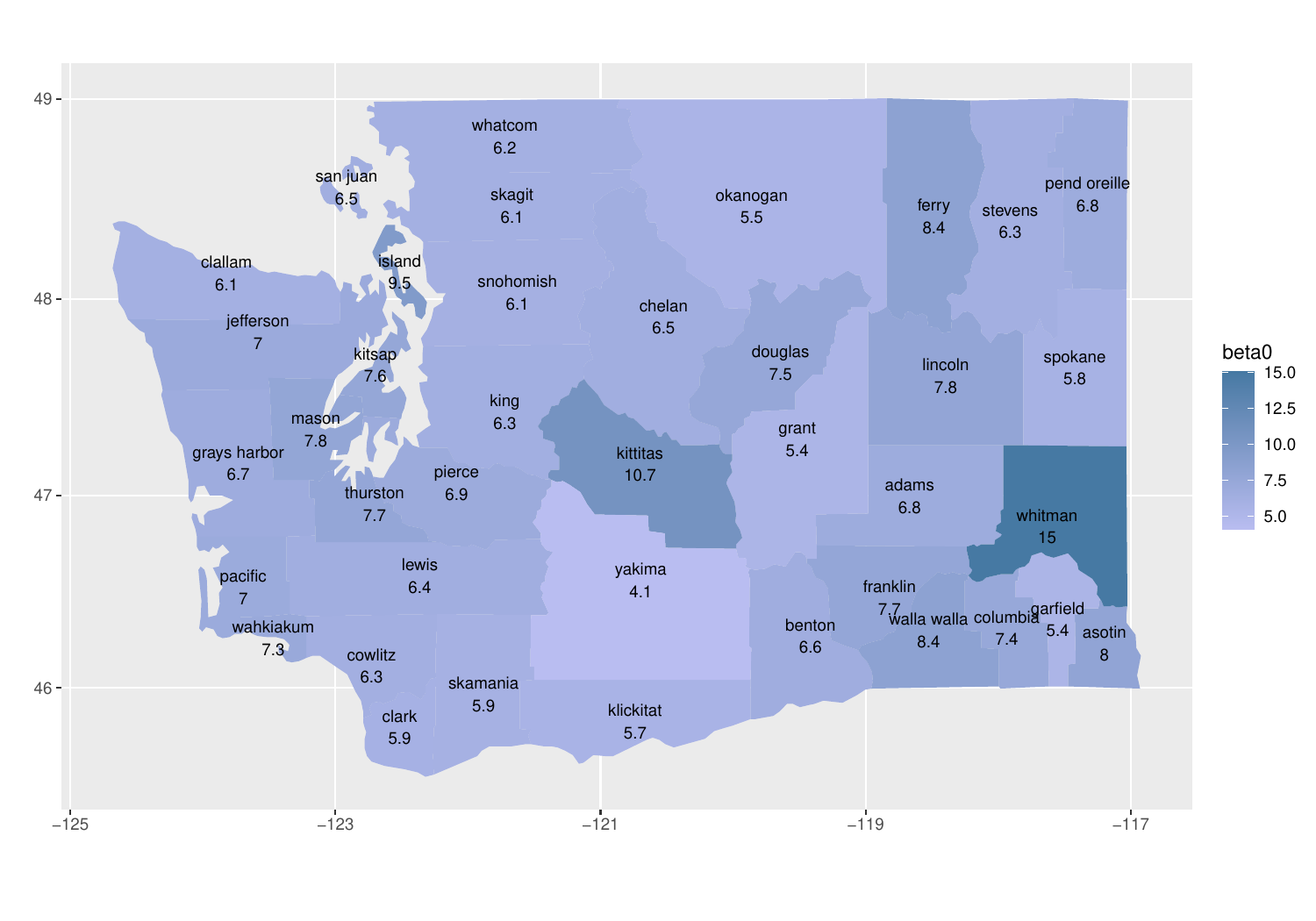}
\end{center}
\caption{\label{fig:mem-beta0-map} \small Map of WA counties showing the random intercepts $\beta_{0,i} \cdot 100$ estimated using the mixed-effect model defined in Equation~\ref{eq:mixedEffectsModelCounties}. }
\end{figure}

The estimated random intercepts $\beta_{0,i}$ (Equations~\ref{eq:mixedEffectsModelCounties} and~\ref{eq:AB}) for each county $i$ can be seen as a map in Figure~\ref{fig:mem-beta0-map}. It is shown as a percentage, i.e. multiplied by 100. It ranges from $4.1\%$ in Yakima to $15\%$ in  Whitman. For the purpose of estimating and projecting age-specific migration in order to project population, we decided to consider the three highest counties, namely Whitman, Kittitas and Island, as outliers and replace their random intercept with the global mean, namely $7\%$.  This is also driven by the fact that the migration in Whitman and Kittitas is mostly attributable to student migration, which our population model is capturing as an external input and thus, it is not modeled. 

\end{document}

%% file: wa_counties_ret_comp.tex
\begin{center}
\begin{tabular}{rlcc}
\hline\hline
\multicolumn{1}{c}{code}&\multicolumn{1}{c}{name}&\multicolumn{1}{c}{IN}&\multicolumn{1}{c}{OUT}\tabularnewline
\hline
$53001$&Adams&--&x\tabularnewline
$53003$&Asotin&x&--\tabularnewline
$53005$&Benton&--&--\tabularnewline
$53007$&Chelan&x&--\tabularnewline
$53009$&Clallam&x&--\tabularnewline
$53011$&Clark&x&--\tabularnewline
$53013$&Columbia&x&--\tabularnewline
$53015$&Cowlitz&x&--\tabularnewline
$53017$&Douglas&x&--\tabularnewline
$53019$&Ferry&x&--\tabularnewline
$53021$&Franklin&--&--\tabularnewline
$53023$&Garfield&x&--\tabularnewline
$53025$&Grant&x&--\tabularnewline
$53027$&Grays Harbor&x&--\tabularnewline
$53029$&Island&x&--\tabularnewline
$53031$&Jefferson&x&--\tabularnewline
$53033$&King&--&x\tabularnewline
$53035$&Kitsap&x&--\tabularnewline
$53037$&Kittitas&--&--\tabularnewline
$53039$&Klickitat&x&--\tabularnewline
$53041$&Lewis&x&--\tabularnewline
$53043$&Lincoln&x&--\tabularnewline
$53045$&Mason&x&--\tabularnewline
$53047$&Okanogan&x&--\tabularnewline
$53049$&Pacific&x&--\tabularnewline
$53051$&Pend Oreille&x&--\tabularnewline
$53053$&Pierce&--&--\tabularnewline
$53055$&San Juan&x&--\tabularnewline
$53057$&Skagit&x&--\tabularnewline
$53059$&Skamania&x&--\tabularnewline
$53061$&Snohomish&--&--\tabularnewline
$53063$&Spokane&x&--\tabularnewline
$53065$&Stevens&x&--\tabularnewline
$53067$&Thurston&x&--\tabularnewline
$53069$&Wahkiakum&x&--\tabularnewline
$53071$&Walla Walla&x&--\tabularnewline
$53073$&Whatcom&x&--\tabularnewline
$53075$&Whitman&--&--\tabularnewline
$53077$&Yakima&x&--\tabularnewline
\hline
\end{tabular}\end{center}